# Laboratory coded aperture imaging experiments: radial hole coded masks and depth-sensitive CZT detectors

J. Hong$^{a*}$, S. V. Vadawale$^{a}$, M. Zhang$^{a}$,

E. C. Bellm$^{a}$, A. Yousef$^{a}$, J. Noss$^{a}$, J. E. Grindlay$^{a}$, and T. Narita$^{b}$

$^{a}$Harvard Smithsonian Center for Astrophysics, 60 Garden St., Cambridge, MA 02138
$^{b}$College of the Holy Cross, 1 College St., Worcester, MA 01610

**ABSTRACT**

The proposed black-hole finder mission EXIST will consist of multiple wide-field hard X-ray coded-aperture telescopes. The high science goals set for the mission require innovations in telescope design. In particular, wide energy band coverage and fine angular resolution require relatively thick coded masks and thick detectors compared to their pixel size, which may introduce mask self-collimation and depth-induced image blurring with conventional design approaches. Previously we proposed relatively simple solutions to these potential problems: radial hole for mask self-collimation and cathode depth sensing detector for image blurring. We have now performed laboratory experiments to explore the potential of these two techniques. The experimental results show that the radial hole mask greatly alleviates mask self-collimation and a ~1 mm resolution depth-sensitive detector scheme can be relatively easily achieved for the large scale required for EXIST.

## 1. INTRODUCTION

The Energetic X-ray Imaging Survey Telescope (EXIST) is proposed as the Black Hole finder mission in the NASA Beyond Einstein program and it was selected for a concept study under the program. EXIST will be a next generation wide-field hard X-ray survey mission, monitoring the full sky every orbit.[1] EXIST consists of multiple wide-field hard X-ray coded-aperture telescopes employing a large area of tungsten mask (~24 m$^2$) and CZT detectors (~6 – 8 m$^2$). Its ambitious scientific goals require innovations in telescope design. The wide energy band coverage (~10 – 600 keV) demands relatively thick mask and detectors. In addition, fine angular resolution (~5′) requires small pixels in the mask and detector. We have noted that the high aspect ratio of the pixel size to thickness may introduce serious imaging issues with conventional design approaches. In the case of the coded mask, the high aspect ratio introduces self-collimation for off-axis photons, and in the case of the detector, it causes depth-induced image blurring for off-axis fields. This can be problematic because the telescopes in EXIST need to cover a large instantaneous field of view (FoV) to achieve the scientific goals.

Previously we proposed simple solutions for these problems.[2,3] For mask design, through 1D mask simulations, we demonstrated the self-collimation can be drastically reduced by radially-oriented holes in the mask rather than conventional vertical holes.[2] For detector design, an obvious solution for image blurring is acquiring depth information of each event, but this should be done over a large area of detectors. To do so, we introduced a simple depth sensing scheme that can be easily modularized and tightly packaged to a large area.[3]

In order to explore the potential of both techniques, we have performed laboratory experiments. The result confirms that indeed these techniques can resolve the potential imaging issues for wide-field hard X-ray coded-aperture telescopes. First, through simple imaging experiments, we compare the X-ray throughput of a radial hole mask with a conventional vertical hole mask and study their imaging performance. Second, we present the initial test results of our prototype large-area depth-sensing module and we establish the depth sensitivity of the detector experimentally through an absolute depth calibration setup.

---
$^{*}$ Send correspondence to J. Hong (jaesub@head.cfa.harvard.edu)

## 2. RADIAL HOLE MASK

Coded aperture imaging is a technique to localize or image a source from decoding a shadow-gram recoded on the detector, which is cast by the source through the mask with a predetermined pattern. The technique is popular in wide-field hard X-ray imaging where the current focusing telescope technology is not practical. Conventional coded-aperture telescopes employ a relatively thin mask, compared to its pixel size. By doing so, there is no collimation and shadow-grams recorded in the detector are a simple superposition of different parts of mask patterns, which can be easily decoded.

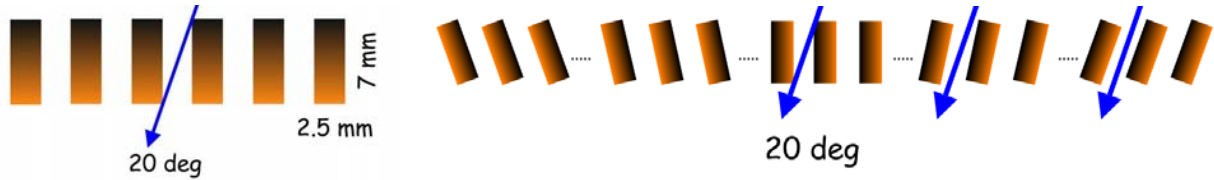

**Figure 1**. *Left*: Self-collimation by uniform mask holes of 2.5mm wide in coded mask of thickness 7mm.[1] *Right:* Radial mask hole concept (left-most ray is outside FoV).[1]

In the case of EXIST, the mask needs to be about 7 mm thick even with tungsten, which stops about 80% of 600 keV X-rays, and the mask pixel should be 2.5 mm to achieve 5′ angular resolution for 1.5 m distance between mask and detector. The high aspect ratio of mask thickness to its pixel size may introduce severe self-collimation for off-axis photons, as illustrated in Figure 1 (left). The mask with conventional vertical holes for open elements will severely collimate photons incoming at off-axis angles greater than $10^o$ in the figure. In addition, the shadow-gram in the detector will not simply follow the mask pattern even for a point source due to collimation, depending on the off-axis angle.

A simple solution to the problem is arranging open elements in the mask radially so that the overall X-ray transmission through the mask can be uniform across the field (Figure 1 right). Previously we demonstrated the X-ray transmission performance of such radial hole masks by 1D mask simulations.[2] In order to verify the simulation results and to explore various imaging aspects of a radial hole mask, we have performed a series of laboratory imaging experiments with a radial hole mask and a vertical hole mask.

### 2.1. Setup for imaging experiment

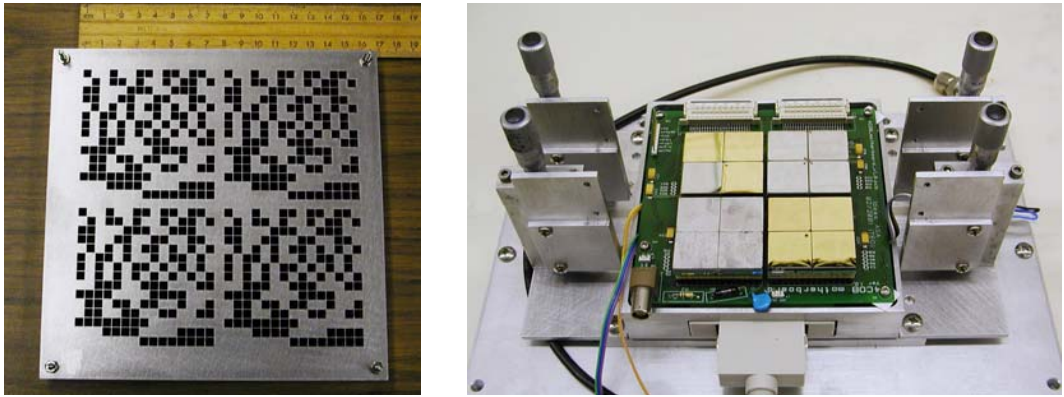

**Figure 2.** *Left*: Picture of the radial hole mask consisting of 2x2 cycles of the basic 16x16 pixel URA pattern. *Right*: the CZT detector module with 4 DCAs (16 crystals) with a 1 pixel gap between DCAs.

For imaging experiments, we have made two masks; one radial hole mask and one vertical hole mask. Each mask consists of seven layers of a stainless steel plate and the mask pattern is laid out by the laser etching technique. Each plate is 2 mm thick, so the final mask is 14 mm thick in total. The mask pixel is 5mm x5mm in the middle layer with ~0.5mm thick grid pattern to support isolated opaque elements. The aspect ratio of thickness to pixel size is equivalent to that of the base-line mask design for EXIST, although the pixel size and mask thickness are double. The mask pattern is 2x2 uniformly redundant array (URA) generated by the perfect binary array method.[4] The fraction of the open elements in the mask is 47% and the overall mask pattern covers 16x16 cm$^2$ with 32x32 pixels. For the vertical hole mask, all seven layers have an identical mask pattern, which is the same pattern of the middle layer of the radial hole mask. For the radial hole mask, the pixel pitch in each layer is off-set by ~0.02 mm from one layer to another so that the orientation of the holes at the edge of the mask is about 10º off from the vertical direction.

For the detector, we use our third generation of the CZT detector system (CZT3) shown in Figure 2.[3,5] The system can accommodate four smaller detector module called DCA (Detector Crystal Array), which consists of 2x2 CZT crystals bonded on a single coupling board. Underneath the board, there are two XAIM 3.2 or 3.3 ASICs to handle 256 anode signals from the four crystals. The dimension of each crystal is 19.44x19.44 mm$^2$ wide, 5 mm thick, and the anode pixel pitch is 2.46 mm. The total detector area of the 16 crystals in the CZT detector system is 8x8 cm$^2$, which consists of 32x32 pixels. Among 16 crystals, 7 crystals have Au/Au contact, and 5 have Pt/Pt contact, and 4 have In/In contact. There is a 1 pixel gap between the DCA modules. The detector was biased at -600V during the imaging experiments.

The overall setup for the imaging experiment is illustrated in Figure 3. For a given set of the instruments, the setup is determined by four parameters; the incident radiation angle ($\theta$), the relative orientation of the telescope with respect to incoming radiation ($\varphi$), the distance between the radiation source and the detector ($D_s$), and the distance between the detector and the mask ($D_m$). For a given setup, we performed a series of measurements with different mask setups with time order: no mask, vertical hole mask, no mask, radial hole mask, and no mask. No mask data is used to compensate for temporal and spatial variation of the detector efficiency and is essential to provide a fair performance comparison of the vertical and radial hole mask. No mask data is also used to remove the structural shadows (e.g. from the high voltage (HV) bias board) other than from the coded masks.

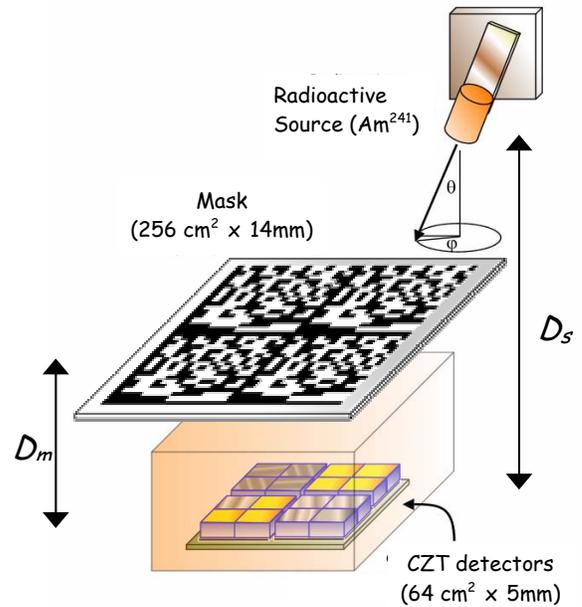

The source holder allows $\theta$ can be any angle from 0º to 15º in 1-degree steps. The telescope orientation $\varphi$ can be simply set by rotating the telescope. The relative position and orientation of the telescope and source is set by a laser mounted on the source holder and the removable guide mark on the mask. $D_m$ can be easily adjusted from 20 cm to 40 cm, depending on off-axis angles to explore. $D_s$ is limited by the laboratory space (~3 – 10 m). The larger $D_s$ is the smaller the magnification factor ($D_m/D_s$) for the shadow pattern is. We use an Am$^{241}$ radioactive source (60 keV) for the X-ray source. We report the result from the initial experiments using the following configurations.

- $\theta = 0º, 2º, 4º, 6º, 8º,$ and $10º$
- $\varphi = 0º$ and $45º$
- $D_s = 2.7$ m
- $D_m = 34$ cm.

**Figure 3** Experimental setup for imaging with vertical vs. radial hole mask

## 2.2. Throughput Comparison

Figure 4 shows the measured X-ray transmission factor through two masks as a function of off-axis angle. The transmission factor is normalized by no mask data with an appropriate factor compensating for temporal efficiency variations. In the figure, the (blue) triangles represent the throughput for the radial hole mask and the (red) circles for the vertical hole mask. The left panel in the figure is for $\varphi=0º$, where the radiation beam is aligned with the orientation of the mask pattern, and the right panel is for $\varphi=45º$, where the beam is incoming diagonally to the mask pattern. In both cases, the radial hole mask has a more uniform response across the FoV, as expected from the previous 1D simulations.[2]

In addition, note that the X-ray throughput enhancement at large off-axis by the radial hole mask is more enhanced at $\varphi=45°$ than at $\varphi=0°$.

The statistical errors in the plot are smaller than the size of the symbols. Therefore, the somewhat unsmooth trend (e.g. at $\theta=6°$ and $\varphi=45°$) must be from the systematic errors. For instance, the gap in the detector and magnification factor due to finite source distance may introduce some variation of the overall effective fraction of open mask elements for a given setup. The relative uncertainty of $\theta$ and $\varphi$ is about $0.2°$, which is not big enough to explain the fluctuations.

The left panel in the figure ($\varphi=0°$) is close in configuration to the 1D mask simulations but 2D mask simulations are required to explain the experimental results completely, in particular for the case of $\varphi=45°$. The off-axis X-ray throughput enhancement by the radial hole mask is somewhat smaller than what we can naively expect from 1D simulations in a similar configuration. Unlike celestial sources that shine parallel radiation to the telescope, the radioactive source at the finite distance will cast a diverging beam on the mask. As a result, the transmission through the radial hole mask will be smaller than what 1D simulations predicts using parallel beams.

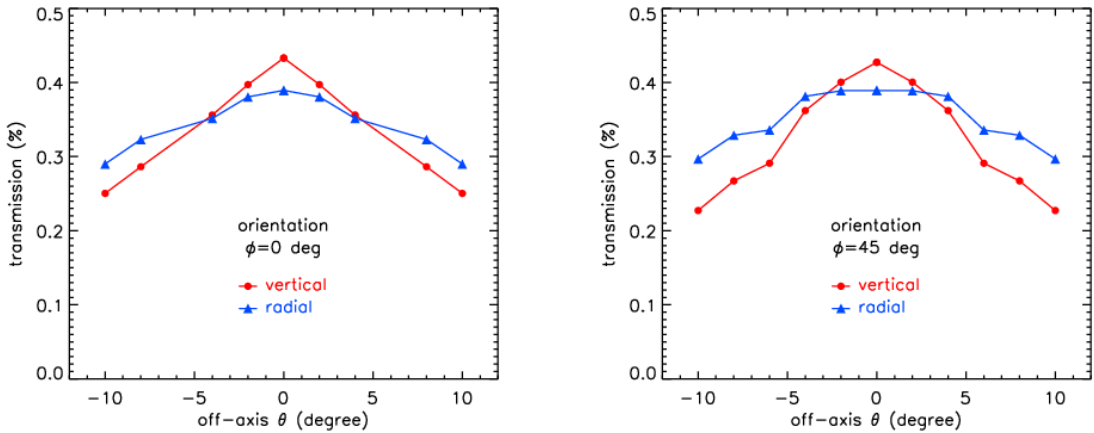

**Figure 4.** X-ray throughput comparison. *Left*: $\varphi=0°$. *Right*: $\varphi=45°$. The (blue) triangles are the results from the radial hole mask and the (red) circles are for the vertical hole mask. The radial hole mask shows more uniform response over the FoV. Note that the result for $\theta=6°$ at $\varphi=0°$ is omitted due to electronics noise.

## 2.3. Imaging Performance

The off-axis X-ray throughput enhancement is an important factor in designing wide-field imaging telescopes. However, in order to understand and thus optimize the imaging performance of the system, we need to understand the imaging artifacts that may be caused by our non-ideal mask, telescope setup, and detector array. Regardless of hole orientations, we expect some imaging artifacts will arise from thick masks and these artifacts will degrade the imaging performance of the system, compared to ideal coded-aperture imaging. The experimental results provide an excellent opportunity to study various realistic imaging artifacts of the coded-aperture system, including ones caused by the finite thickness of the mask.

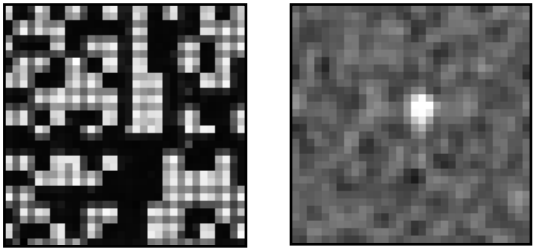

**Figure 5.** Imaging with the radial hole mask. *Left*: a detector shadow-gram for $\theta=0°$, $\varphi=45°$. *Right*: the reconstructed image by cross-correlating the shadow-gram with mask pattern.

Figure 5 shows the detector shadow-gram generated by the radial hole mask at $\theta=0°$, $\varphi=45°$. The shadow-gram is corrected for the spatial variation of detector efficiency using no mask data. Due to the magnification (~12.3%) caused by the finite source distance, the shadow pattern is only ~ 77% of the 1x1 cycle of the URA pattern. Figure 5 also shows the reconstructed image by simply cross-correlating the shadow-gram with the mask pattern. The reconstructed image shows a strong point source, indicating the presence of the radiation source, but the signal-to-noise ratio (SNR) of the point source in the image is smaller (105) than what is expected from an ideal system governed only by Poisson statistics (3440; see Table 1). In fact, under the same configuration, the imaging experiment with the vertical mask produces a very similar detector shadow-gram and the image, which is not distinguishable by eye from the above result, except for the overall intensity level.

In order to understand the imaging result, we list the systematic errors of the system that separate our experiments from an ideal coded-aperture imaging (in the order approaching from the detector to the source).

  (a) The detector efficiency variation; both temporal (~5–10 %) and spatial (~15%).

  (b) The detector gap between DCA modules (1 pixel).

  (c) Structural shadows: the HV bias board with the pogo-pins (up to a factor of ~5 variations in detection efficiency).

  (d) Non-ideal mask; the thickness of the mask, the finite transmission factor of the stainless steel.

  (e) Magnification of the shadow gram of the mask pattern due to the finite source distance (~12.3%).

In order to remove the systematic noise due to (a) and (c), we generate the detector shadow-gram from the ratio of two data sets taken with and without the mask using an appropriate normalization factor. The practice can be justified from the fact that the same procedure applied for two different sets of no-mask data produces a Poisson noise dominant distribution.

Among the rest of the systematics, ultimately what we are interested in is the level of the systematic noise caused by (d). The systematic noise due to (b) and (e) are caused by the limitation in the current imaging system. In order to understand these sources of systematic noise, particularly ones caused by (d), either we need to perform experiments relatively free of (b) and (e) or need to conduct complete 3D Monte-Carlo simulations taking into account of the 3D mask geometry with the proper X-ray transmission physics. Nevertheless, as shown below, we can estimate the effect of the imaging artifacts from (d) using relatively simple semi-analytic simulations. The result indicates that (e) is the dominant source of the residual systematics among the three components.

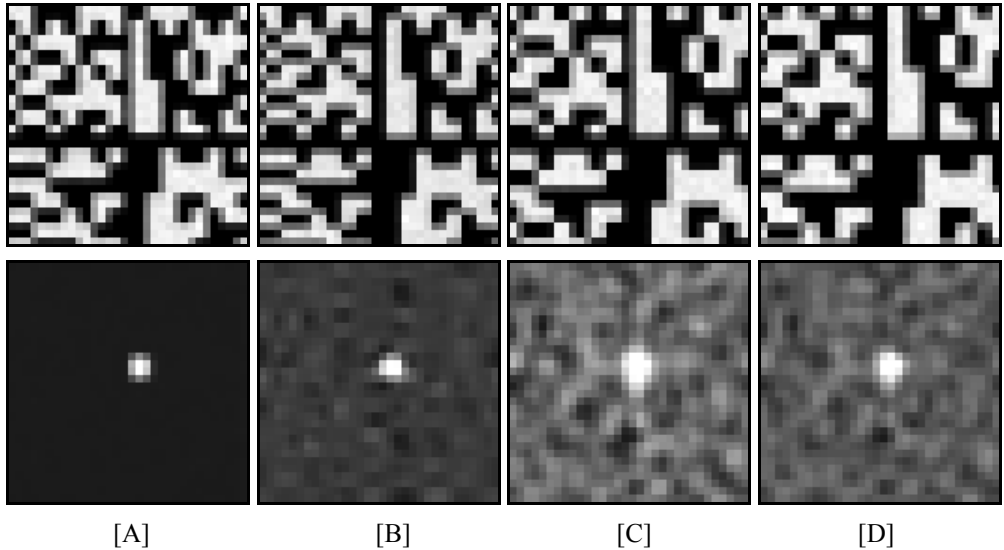

[A]　　　　　　　　[B]　　　　　　　　[C]　　　　　　　　[D]

**Figure 6.** Imaging simulation. *Top*: simulated detector shadow-grams for $\theta=0^o$, $\varphi=45^o$ with various artifacts. *Bottom*: corresponding shadow-gram by simply cross-correlation. From left to right, the simulation is for an ideal case [A], a case with a detector gap (0.8 pixel) [B], a case with magnification (13.25%) [C], and a case with both detector gap and magnification [D]. Note the similarity of the simulated image at the bottom right with the image in Figure 5.

Figure 6 shows a series of the simulated detector shadow-grams and the reconstructed sky image for various artifacts. From left to right, the simulation is for an ideal system [A], a system with the detector gap (0.8 pixel) [B], a system with a magnification factor (13.25%) [C] and a system with both the detector gap and the magnification [D]. All the simulations are done for an infinitesimally thin mask with the infinite opacity for opaque elements, so that these results can be relatively easily derived.

One can easily notice that the simulation image [D] is almost indistinguishable with the real image in Figure 5, which implies that the systematics due to (b) and (e) may be sufficient to explain the systematic noise in the real image. Although the simulation image [D] appears to be identical to the real image by eye, there is still some additional systematic noise in the real image, which is likely caused by the non-ideal mask geometry (d). In fact, the detector gap (0.8 pixel) and the magnification factor (13.25%) in the simulations are chosen to match the simulated image [D] with the real image, so they are slightly different from the true value (1 pixel gap and ~12.3% magnification), which indicates the additional sources of systematic noise.

Table 1 breaks down the various statistics of the simulations and the real data. The table contains four simulation results in Figure 6 and two experimental results with the radial and vertical hole mask for $\theta=0^\circ$, $\varphi =45^\circ$. First, the SNR values indicate that the experimental images are roughly a factor of 30 worse than the ideal case [A]. The same conclusion can be drawn from the noise RMS. As far as the SNR or the noise RMS goes, the simulation [D] produces a result very similar to the actual data, which is expected from the similarity in their images.

In order to investigate this apparent similarity of the images in detail, we calculated the $\chi^2$ and RMS of the difference of two images as shown in the table. $\chi^2 \sim 1$ or RMS ~ 0.5 means the two images are identical within the statistical noise. In the case of comparisons with the radial hole mask data, simply adding the detector gap [B] in the simulation ($\chi^2=1700$) generates an image even more different from the real image than the ideal case [A] does ($\chi^2=1100$). The magnification alone [C] ($\chi^2=430$) or the magnification and the detector gap together [D] ($\chi^2=101$) reduces the gap between the simulated image and the real image. However, $\chi^2 \gg 1$ indicates that there is still some residual imaging artifacts in the real image beside the detector gap and the magnification.

The real image with the vertical hole mask is more similar to the simulation image [D] ($\chi^2=62.2$) than the image with the radial hole mask ($\chi^2=101$) is. This is somewhat expected particularly for the experiments with normal incident ($\theta=0$), diverging beams on the mask, where the systematics caused by mask thickness stands out more with radial holes. Note that $\chi^2$ between two real images with the radial and vertical hole mask is smaller (17.1) than that between the simulation image and the real image. The last column in the table shows the comparison with another simulated result under the condition [D] for a consistency check.

Naively judging from the RMS values of the image difference between the simulation [D] and the real data (RMS=3.98 or 3.14), the degradation of image quality by the non-ideal mask is roughly a factor of 8 or less (pure Poisson RMS~0.5 here). Although it may take full 3D simulations to understand the imaging artifacts completely due to non-linear effects of different systematic noises (e.g. simply adding the detector gap in the simulation makes the simulations less accurate than an ideal case does), the above degradation factor is quite interesting. Since the required dynamic range of EXIST is very large (~$10^4 - 10^5$), losing a factor of 10 (at most) in the imaging performance simply from the mask appears to be critical. However, such a mask-induced degradation factor likely depends inversely on (the square root of) the number of the mask elements (e.g. the coding noise by a random mask pattern compared to an ideal URA pattern is directly related to the square root of the number of the mask elements) and considering the ~300x300 mask elements for EXIST sub-telescopes, the resulting imaging degradation by this type of non-ideal masks for EXIST telescopes may be quite negligible. Full 3D simulations are planned.

**Table 1** Run-down statistics of simulated image and experiments and their comparison.

| Data Type | | SNR | Noise RMS* | vs. Radial | | vs. Vertical | | vs. Sim [D] | |
|---|---|---|---|---|---|---|---|---|---|
| | | | | $\chi^2$ | RMS* | $\chi^2$ | RMS* | $\chi^2$ | RMS* |
| Simulations | [A] Ideal case | **3440** | **0.425** | **1100** | **13.5** | **1060** | **13.3** | **981** | **12.7** |
| | [B] detector gap (b) | 217 | 6.35 | 1700 | 16.5 | 1640 | 16.2 | 1570 | 15.9 |
| | [C] magnification (e) | 108 | 13.2 | 430 | 8.25 | 380 | 7.78 | 308 | 7.03 |
| | [D] both (b) and (e) | **128** | **11.2** | **101** | **3.98** | **62.2** | **3.14** | **1.06** | **0.600** |
| Real data | Radial hole mask | 105 | 12.1 | - | | 17.1 | 1.64 | 101 | 3.98 |
| | Vertical hole mask | 110 | 12.9 | 17.1 | 1.64 | - | | 62.2 | 3.14 |

RMS*: in an arbitrary unit, meaningful in comparison

# 3. DEPTH SENSITIVE CZT DETECTOR

The detector design for EXIST is also challenging. EXIST will employ CZT detectors for their various advantages such as high spectral and positional sensitivities.[1,3] As for the mask, the broad energy band coverage (10 – 600 keV) requires thick detectors (~ 5 – 10 mm), and high angular resolution (~5′) to avoid source confusion requires small detector pixels (1.25 mm pitch for 1.5 m separation of masks and detectors). The high sensitivity (~ 0.05 mCrab below ~100 keV) and long duty cycle (≥20%) requirements call for a large area detector (~ 8 $m^2$) and wide field of view (170$^o$ × 75$^o$) for full-sky imaging each orbit.

In coded-aperture imaging, when reconstructing sky images from data taken by detectors with a relatively large ratio of thickness to pixel diameter the lack of depth information for X-ray interactions is effectively translated into inaccurate pixel identification for X-rays coming at a far off-axis angle. The resulting ambiguity in pixel location on the imaging plane degrades the angular resolution of the telescope. Therefore, depth sensing is necessary for thick, small pixel detectors for wide-field imaging.

Besides restoring angular resolution, depth sensing can improve energy resolution by allowing correction for depth-dependent charge collection efficiency or other anomalies. Depth sensing also enhances background rejection by singling out certain types of background events such as shield leakage and internal background. With depth sensing, one can operate detectors in Compton telescope mode. Finally depth sensing can improve the polarization sensitivity of detectors if there is also multi-site (multi-pixel) event readout capability.

The technique we explore is called cathode depth sensing, which is acquiring depth information of X-ray interactions by simultaneous measurements of cathode and anode signals from the interactions in a pixellated detector.[7,8] In our detector, the anode contact is pixellated and the cathode side faces X-ray sources due to the complex electronics readout systems closely attached to the anode side. For a given X-ray interaction in the detector, the signal gathered by an anode pixel is proportional to the charge induced by the interaction, while the cathode side collects a signal that is proportional to the induced charge times the depth of the interaction. Therefore, the ratio of two signals is roughly proportional to the depth of the interaction.

## 3.1. Detector setup

Previously we have introduced a cathode depth sensing scheme using an electronics board that is vertically mountable on the cathode surface as illustrated in Figure 7.[3] The vertically mountable cathode read-out board offers two advantages for EXIST. First, it can be easily upgraded to a modular system that can be closely tiled into a large area. Second, the board can be used as a low energy collimator, which may be required due to a relatively high population of bright soft X-ray sources in the sky.

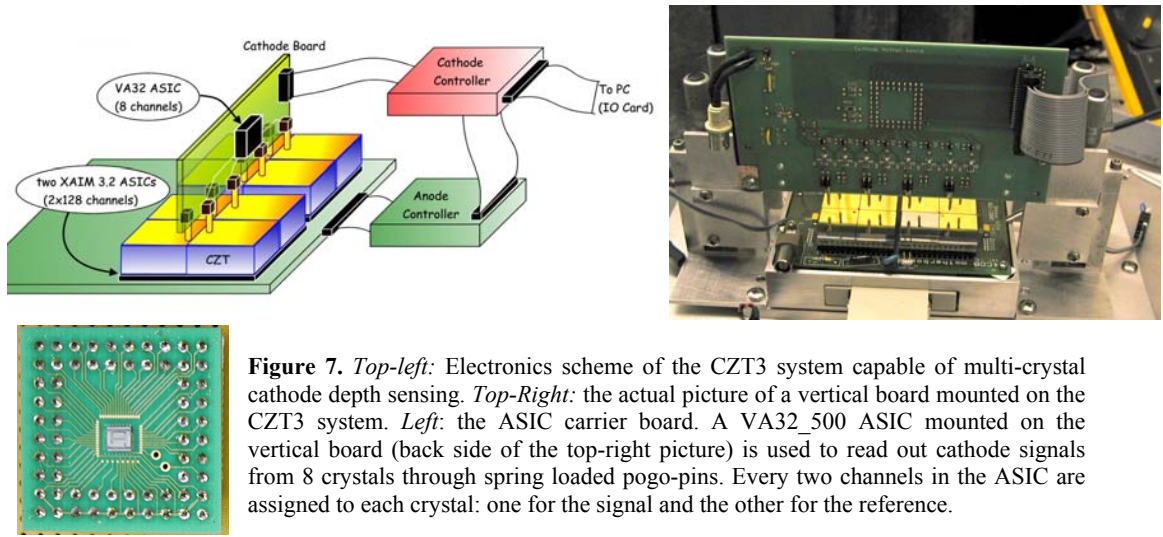

**Figure 7.** *Top-left:* Electronics scheme of the CZT3 system capable of multi-crystal cathode depth sensing. *Top-Right:* the actual picture of a vertical board mounted on the CZT3 system. *Left*: the ASIC carrier board. A VA32_500 ASIC mounted on the vertical board (back side of the top-right picture) is used to read out cathode signals from 8 crystals through spring loaded pogo-pins. Every two channels in the ASIC are assigned to each crystal: one for the signal and the other for the reference.

In the current prototype cathode board, the cathode signal from each crystal is read out through a spring-loaded pogo-pin directly contacting the cathode surface of the crystal. The cathode signal is amplified in an IDEAS VA32_500

ASIC, which can process 32 channels in parallel. The ASIC is mounted on the vertical cathode board through a small carrier board. Every two channels in the ASIC are assigned to each crystal: one for the signal and the other for the reference for the differential output to cancel out electronics noise. In the current configuration, the VA32_500 ASIC handles signals from 8 crystals, but one VA32_500 ASIC can handle as many as 16 crystals in this scheme.

In this prototype cathode board, each component is somewhat spaciously placed for test and modification. The length of the input trace line for the cathode signal varies from ~5 to 10 cm, depending on the channels. According to ASIC testings with external pulse signals, all 16 channels we use in the ASIC are functioning properly. However, channels with a long input trace line tend to exhibit a lower gain than other channels with a short input line. We found the two channels that connect to the pogo-pins in the middle of the board (the shortest trace line) show the highest gain (x~50) and others vary from x10 to x40. All the reference channels with no input trace line show a high gain (x~50) as well. We also found the input lines shielded by surrounding ground lines tend to show a high gain (half of the signal channels are designed in this way for testing).

The optimal HV bias for anode signals is expected around -900V for these crystals.[6] Here the initial data were taken at HV~ -600V due to a HV-induced noise in the cathode electronics. The overall cathode read-out system performs properly, but the system requires another revision to take advantage of its potential. The resolution of the cathode signal from the current prototype board is not as good as that with our earlier signal channel system using an AmpTeK preamplifier.[3]

### 3.1. Absolute depth calibration

For depth sensing experiments, we also developed a simple source mounting scheme that allows the absolute depth calibration for the detector, as shown in Figure 8. In this setup, we generate a finely collimated fan beam of radiation using a tungsten carbide slit. The radiation slices through the detector, and for a given pixel, the path of the radiation is confined to a narrow range of the depth. Pixel identification directly indicates the depth of events, which can be used to calibrate the signals. The slit size can be easily adjusted in steps of ~1 mil using a thin Al shim between two tungsten carbide bars. This setup allows an easy calibration of a large area of detectors and the beam alignment with low energy radiations is relatively easy, compared to the case using a horizontal beam.[3]

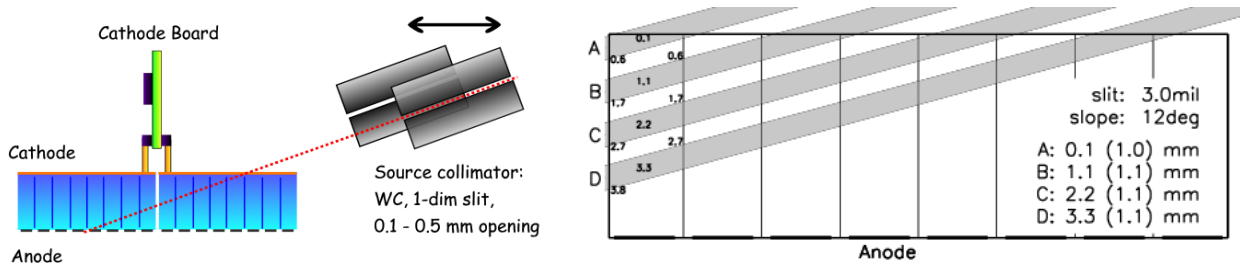

**Figure 8.** *Left*: The absolute depth calibration scheme. Two 2×2×5 cm³ tungsten carbide bars are used to collimate the radiation from a source (not shown) into a very narrow off-axis fan beam. The collimator mounting stage can set the beam elevation from 10° to 25° or 0° (horizontal beam) and the whole stage is mounted on a translation micrometer that can control the precise location of the irradiation point on the crystals. *Right:* the radiation path in a crystal for the example data in Section 3.3. Four measurement configurations, labeled in A, B, C and D, are set with the 12° beam elevation and the 3 mil slit opening. The numbers next to the labels are the center of the radiation depth for edge pixels and the spread of the radiation depth in the pixel (in the parenthesis).

### 3.3. Initial Measurement and Depth sensitivity

We present some of the initial measurement results and derive the experimental depth sensitivity. The right panel in Figure 8 illustrates the radiation path in a crystal for the initial data set, assuming a perfect collimation from tungsten carbide with 100% opacity. The slit size is 3 mil, and the source distance is about 20 cm from the detector, and the radiation is incoming at 12° elevation. The right panel in the figure also shows the expected depth of the center of the beam spread in the pixels at the edge row for four different configurations, which are labeled in A, B, C and D. The radiation depth change between two neighboring setups is about 1.0 - 1.1 mm. The depth spread in a pixel for each set is also about 1.0 - 1.1 mm. In the case of A, the depth center would be somewhat deeper than 0.1 mm (so the spread is smaller than 1.0 mm) since the crystal is only 5.0 mm tall. Therefore, the depth difference of A and B would be somewhat smaller than 1.0 mm. It should be noted that the outmost pixel at the edge of the crystal is about 2.34 mm wide and inner pixels are about 2.46 mm wide. Therefore, the beam spread in the inner pixels is about 5% wider than that in the edge pixels.

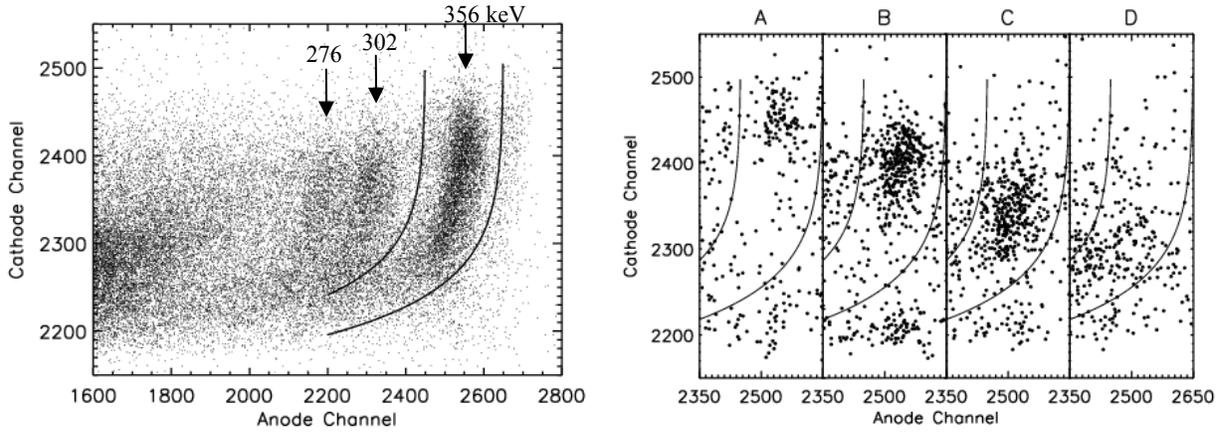

**Figure 9.** *Left*: cathode vs. anode signal for six edge pixels of a Pt/Pt IMARAD crystal from all the measurements. Three clustering of the events indicated by the arrows are from three high energy lines of the Ba[133] source. The two solid curves enclose only the 356 keV events. The relatively broad distribution of the anode signal for a given energy is mainly due to over-plotting results from six anode pixels that have a slightly different gain. *Right:* the same as the left panel but showing each measurement of the four radiation configurations separately (labeled in A, B, C, and D in the Figure 8). As the depth of the events changes, the event cluster in this phase space moves down, indicating the depth sensitivity of the system.

The left panel in Figure 9 shows cathode vs. anode signals generated by a Ba[133] radioactive source. The plot is generated by adding all the data together (including the above four configurations) for the six pixels at the left edge (see the right panel in Figure 8) except for two corner pixels, which show a somewhat different anode gain than what the six pixels have. The Ba[133] source generates four high energy lines in the anode signal range of the plot; 276, 302, 356, and 383 keV. In the plot, the 356 keV line is most prominent, which is surrounded by the two solid curves. We focus on the 356 keV events here. By selecting events that fall between the two solid curves, one can select the 356 keV events exclusively. The solid curves are based on the empirical formula that describes the relation of the cathode and anode signals in the CZT detector.[3] We can perform the same selection procedure for individual sets of the data. The right panel in Figure 9 shows the cathode vs anode signals for each of four data sets corresponding to the four configurations on the right panel of Figure 8. Note that for a given fixed energy, the cathode signal is directly related to the depth of the events. One can easily see that there is a distinct location of the data cluster in the cathode vs. anode phase space, depending on the depth of the events. Due to the noise in the experimental setup, the scatter in Figure 9 is larger than in our single-channel setup.[3] This could be easily improved in a next-generation system.

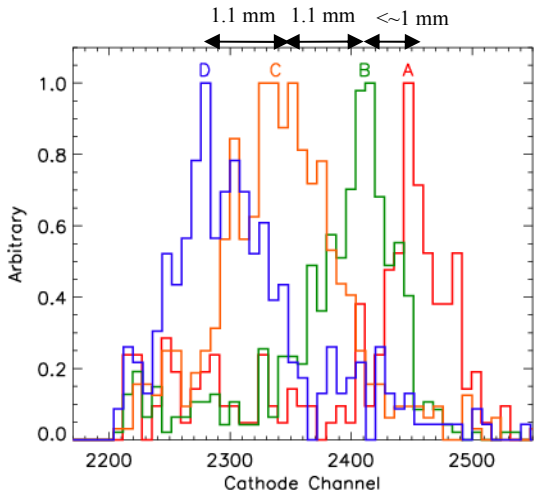

**Figure 10.** Cathode pulse height histograms of the 356 keV events for each of four measurements on the right panel in Figure 8 and 9. The histograms are renormalized to have the identical maximum height for easy interpretation. The three left histograms A, B and C are well resolved, indicating ~1 mm or less depth sensitivity near the cathode surface.

Figure 10 shows the cathode pulse height histograms of the data selected by the solid curves in Figure 9 from the four measurements. Histograms A and B overlap near the half maximum, indicating the depth of the events (<1.0mm) is well resolved. In the case of histograms B and C, they are also well resolved, but for C and D, two histograms overlap around ¾ of the maximum, indicating the depth sensitivity is somewhat worse than 1.0 mm. Note that the distributions are broadened by the electronics noise and, particularly by the finite beam spread (~1.0mm) in the pixel. Figure 10 indicates the sub-mm depth sensitivity holds near the cathode side and it gradually degrades to ~1.5 mm at ~3 or 4 mm deep from the cathode surface. Considering that this is the result from the first round of the cathode electronics design using an ASIC, it is quite possible to achieve the mm-level depth sensitivity with the current scheme. Note that ~1.5 – 2 mm level depth sensitivity is sufficient to resolve the image blurring in EXIST, for which the anode pixel pitch would be 1.2mm.

Finally, the detection efficiency as well as the depth sensitivity drops significantly below ~3 or 4 mm deep from the cathode side. Both efficiency and depth sensitivity near the anode side is greatly depending on the HV bias, and they are expected to improve at the known optimal HV bias (~ -900V) for these crystals.

## 5. CONCLUSION

EXIST will be a next generation wide-field hard X-ray coded-aperture telescopes. Its ambitious scientific goals require innovative instrument design. Potential imaging problems such as mask self-collimation and depth-induced image blurring may occur with conventional design approaches. Here we have experimentally demonstrated that such problems can be easily avoided by relatively two simple techniques - radial hole coded-mask imaging and depth-sensitive detectors.

## ACKNOWLEDGMENTS

This work is supported in part by NASA grant NAG5-5279 and NAG5-5396.